\documentclass[prb,onecolumn,12pt]{revtex4-1}
\usepackage{epsfig}
\usepackage{amssymb}
\usepackage{threeparttable}
\usepackage{graphicx}
\usepackage{natbib}
\usepackage{longtable}
\usepackage{txfonts}
\usepackage{color}

\begin{document}
\title{The anisotropic ultrahigh hole mobility in strain-engineering two-dimensional penta-SiC$_2$ }
\author{Yuanfeng Xu$^1$, Zeyu Ning$^1$, Hao Zhang$^{1,3,*}$, Gang Ni$^1$,  Hezhu Shao$^2$, Bo Peng$^1$, Xiangchao Zhang$^1$, Xiaoying He$^1$, Yongyuan Zhu$^3$ and Heyuan Zhu$^1$}
\affiliation{$^1$Department of Optical Science and Engineering and Key Laboratory of Micro and Nano Photonic Structures (Ministry of Education), Fudan University, Shanghai 200433, China\\
$^2$Ningbo Institute of Materials Technology and Engineering, Chinese Academy of Sciences, Ningbo 315201, China\\
$^3$Nanjing University, National Laboratory of Solid State Microstructure, Nanjing 210093, China}

\begin{abstract}
Using the first-principles calculations based on density functional theory, we systematically investigate the strain-engineering (tensile and compressive strain) electronic, mechanical and transport properties of monolayer penta-SiC$_2$. By applying an in-plane tensile or compressive strain, it is easy to modulate the electronic band structure of monolayer penta-SiC$_2$, which subsequently changes the effective mass of carriers. Furthermore, the obtained electronic properties are predicted to change from indirectly semiconducting to metallic. More interestingly, at room temperature, uniaxial strain can enhance the hole mobility of penta-SiC$_2$ along a particular direction by almost three order in magnitude, $i.e.$ from 2.59 $\times10^3 cm^2/V s$ to 1.14 $\times10^6 cm^2/V s$ (larger than the carrier mobility of graphene, 3.5 $\times10^5 cm^2/V s$), with little influence on the electron mobility. The high carrier mobility of monolayer penta-SiC$_2$ may lead to many potential applications in high-performance electronic and optoelectronic devices.
\end{abstract}

\maketitle
\section{INTRODUCTION}
Two-dimensional (2D) materials have attracted great attention in the fields of nanoscale materials and nanotechnology since the experimental realization of graphene\cite{sciencegraphene, Novo2005, zhang2005}, which possesses remarkable mechanical, electronic and optical properties\cite{alandin2011, Balandin2008, Schedin2007} possibly leading to many novel applications in microelectronic and optoelectronic devices.  A large number of graphene-like structures and other layered structures with different chemical compositions and different crystal structures have been proposed and synthesised since then. Recently, Zhang \textit{et al.}\cite{Zhang2015}, predicted a new 2D material composed of carbon pentagons, $i.e.$ penta-graphene, with mixed $sp^2/sp^3$ orbital hybridization, which is dynamically and mechanically stable, and can withstand temperatures as high as 1000 $K$. Monolayer penta-graphene possesses unusual negative Poisson's ratio, ultrahigh ideal strength outperforming graphene and other interesting electronic properties \cite {Zhang2015,xu2015}, which make it a potential candidate for wide applications in the optoelectronics and photovoltaics devices. 

Many efforts have been devoted on the discovery of new 2D materials with penta strcuture since the successful prediction of penta-graphene. The silicon counterpart of the penta structure was proposed as well, although it is  dynamically instable in the monolayer form\cite{Yierpan2016}. Recently, a novel pentagonal structure composed of carbon and silicon atoms in a $sp^2d^2$ hybridization, $i.e.$ pentagonal silicon dicarbide (penta-SiC$_2$)\cite{Littlewood2015}, is proposed and further confirmed to be dynamically stable and exhibits enhanced similar electronic property compared to penta-graphene.

During the processes of experimental realization of 2D materials, mismatch between 2D material and substrate often results in the formation of crystal distortion due to strain or stress modulation. Experimental measurements of physical properties of 2D materials could be different from those with perfect structures. Therefore it is necessary to investigate the strain-dependence properties of 2D materials comparing experimental results. In addition, strain-engineering (external strain and stress) is an effective way to modulate the electronic, mechanical, optical properties and transport of 2D materials\cite{Feng2012, Conley2013, Yu2016, kang2015}. It has been reported that strain engineering can be used to modulate the electronic properties of MoS$_2$\cite{Cai2014}, black phosphorene\cite{Fei2014},TiS$_3$\cite{Aierken2016} and ReS$_2$\cite{Yu2016}. The strain-engineering method was proved to be effective in increasing the band gap of pengta-SiC$_2$\cite{Littlewood2015} as well. However, further investigation on the carrier mobility of monolayer penta-SiC$_2$ under various strain has not been addressed.

In this work, we systematically investigate electronic, mechanical and transport properties of monolayer penta-SiC$_2$ under tensile and compressive strain by using first-principles calculations. We find that under in-plane uniaxial strain, penta-SiC$_2$ remains an indirect semiconductor until ${22\%}$, however, the uniaxial compressive strain can easily converted it from semiconducting into metallic beyond ${-15\%}$. When applying a biaxial strain, penta-SiC$_2$ can change from semiconductor to metallic material under relatively low compressive strain of ${-8\%}$. In general, the electronic band structure of  penta-SiC$_2$ is sensitive to tensile and compressive strain. Moreover, we have quantitatively demonstrated that the carrier mobility can be enhanced from 2.59 $\times10^3 cm^2/V s$ to 1.14 $\times10^6 cm^2/V s$ by applying a uniaxial compressive strain.

\section{Method and computational details}

The calculations are performed using the Vienna \textit{ab-initio} simulation package (VASP) based on density functional theory \cite{Kresse1996}. The exchange-correlation energy is described by the generalized gradient approximation (GGA) using the Perdew-Burke-Ernzerhof (PBE) functional. The calculationa are carried out by using the projector-augmented-wave pseudo potential method with a plane wave basis set with a kinetic energy cutoff of 500 eV. When optimizing atomic positions, the energy convergence value between two consecutive steps is chosen as 10$^{-5}$ eV and the maximum Hellmann-Feynman force acting on each atom is 10$^{-3}$ eV/\AA. The Monkhorst-Pack scheme is performed for the Brillouin zone integration with k-point meshes of 13$\times $13$\times $1 and 21$\times $21$\times $1 for geometry optimization and self-consistent electronic structure calculations, respectively. To verify the results of the PBE calculations, the electronic structure of penta-SiC$_2$ with zero strain is calculated using hybrid Heyd-Scuseria-Ernzerhof (HSE06) functional\cite{HSE03}. HSE06 improves the precision of band gap by reducing the localization and delocalization errors of PBE and HF functionals. The mixing ratio used in the HSE06 is 25$\%$ for short-range Hartree-Fock exchange. The screening parameter $u$ is set to be 0.4 \AA$^{-1}$.

\begin{figure}
\centering
\includegraphics[width=0.75\linewidth]{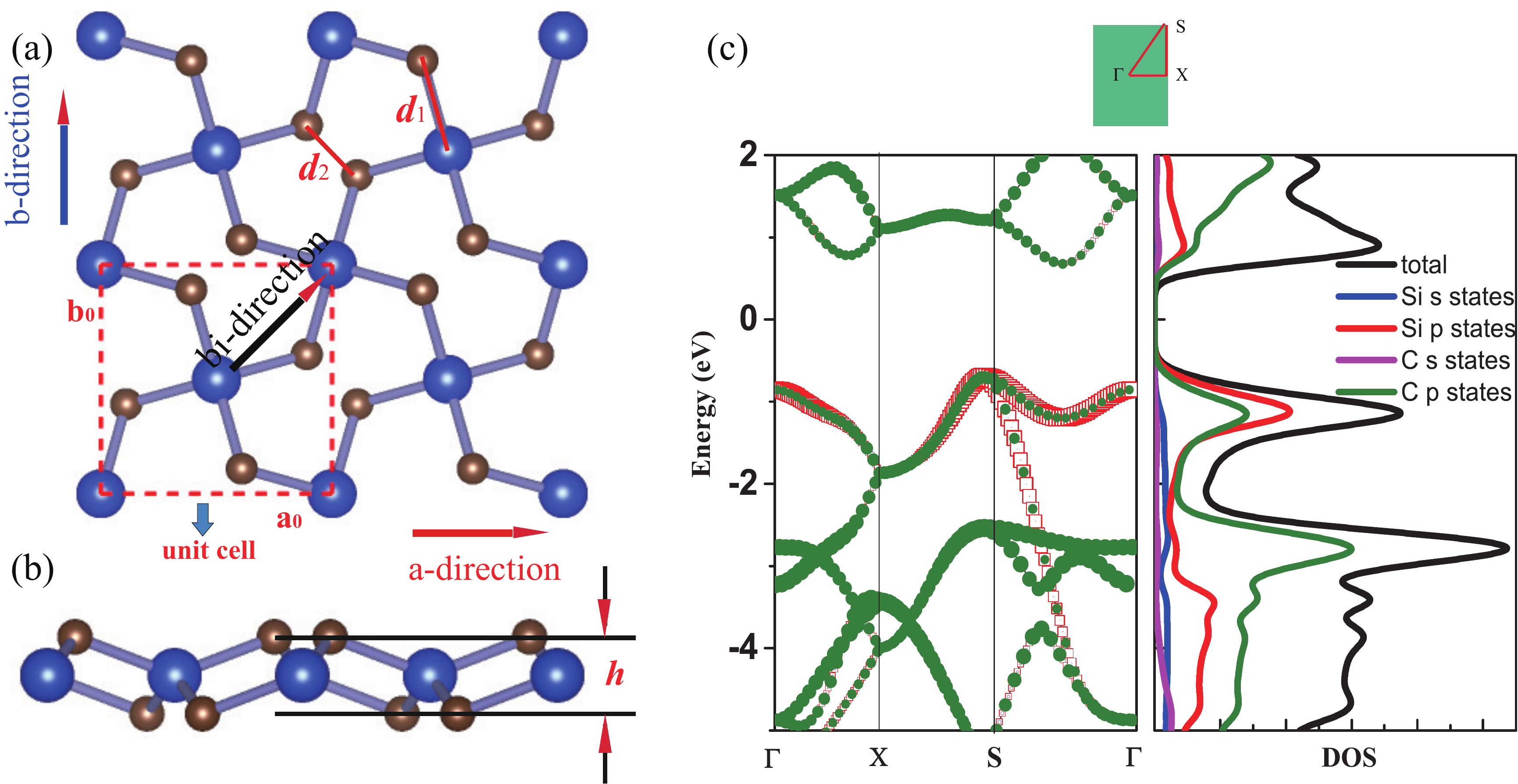}
\caption{(a) Top view (b) side view of the atomic structure of penta-SiC$_2$ monolayer (2$\times $2 supercell). $d_1$ and $d_2$ are the bond length of Si-C and C-C in this pentagonal structures, respectively. $h$ is the buckling distance.  (c) is PBE calculated orbital-projected electronic band structures of monolayer penta-SiC$_2$ along the represents high-symmetry directions of Brillouin zone (BZ) and density of states (DOS) of penta-SiC$_2$.}
\label{structure} 
\end{figure}

\section{Results and discussion}

\subsection{Electronic band structure of penta-SiC$_2$}
The top and side views of the fully optimized structure of penta-SiC$_2$ are shown in Fig.~\ref{structure}(a) and (b), respectively. The unit cell (the red dashed rectangle) contains two silicon atoms and four carbon atoms. The optimized penta-SiC$_2$ has a tetragonal crystal structure with the space group of $p\overline{4}2_1m$. As shown in Fig.~\ref{structure}(a) and (b), all pentagons contain four equivalent Si-C bonds and one C-C bond with three atomic layers (two C atom layers on the top and bottom, and Si atom layer in the middle). As shown in Table.~\ref{tabel-1}, the associated lattice constant of penta-SiC$_2$ is $a$=$b$=4.41 $\AA$ with the buckling height $h$=1.33 $\AA$. The bonding lengths are $d_1$=1.91 $\AA$ and $d_2$=1.36 $\AA$, which are in good agreement with previously theoretical results\cite{Li2015}. To avoid artificial interaction between atom layers, the separation between the layers is set to be 15 $\AA$  for monolayer penta-SiC$_2$. 

The PBE calculations show that, SiC$_2$ is a semiconductor with an indirect band gap of 1.39 eV, and the minimum of the conduction band (VBM) is located at S point, the maximum of the valence band (VBM) is located between S and $\Gamma$ points. The corresponding HSE06 calculation gives a larger band gap of 2.85 eV comparing to the PBE result since PBE always underestimates the value of band gap of semiconductors.  Fig.~\ref{structure}(c) shows the PBE calculations of orbital-projected electronic band structures and density of states (DOS) of penta-SiC$_2$. Analysis on the PDOS (Si-3$s$, 3$p$ and C-2$s$, 2$p$ orbitals) of penta-SiC$_2$ reveals that C-2$p$ and Si-3$p$ orbitals mainly dominate the electronic states near the Fermi level. However, the contributions by the C-2$p$ states to the total DOS is larger than that by Si-3$p$, and in the energy range of -2 to 0 eV for valence band, the Si-3p states contribute larger than that by C-2$p$, indicating the strong hybridization between the Si $3p$ states and C $2p$ states. The analysis on the chemical bonding of pengta-SiC$_2$ based on the DOS is in consistency with the calculated bond lengths, which show an identical length of Si-C bonding. The four identical bond lengths and bonding angles indicate that Si atom connects to the four neighboring C atoms via $sp^2d$ hybridization. Similar analysis on C atoms shows that C connects to the two neighboring Si atoms and one C atom via $sp^3$ hybridization.

\begin{table}
\centering
\caption{Calculated structure properties of 2D penta-SiC$_2$: lattice constant($a$, $b$), buckling distance ($h$), Si-C and C-C bond length ($d_1$, $d_2$). The magnitude of band gap calculated under PBE and HSE06 approximation respectively.}
\begin{tabular}{cccccccc}
\hline
    structure &  lattice constant[\AA] &  $h$[\AA] &  $d_1$[\AA] &  $d_2$[\AA] & $E_g(PBE)[eV]$ & $E_g(HSE)[eV]$ & Young's modulus(GPa) \\
\hline
penta-SiC$_2$ & 4.41 & 1.33 & 1.91 & 1.36 & 1.39 & 2.85 &292 \\
\hline
\end{tabular}
\label{tabel-1}
\end{table}

\subsection{Strain-engineering mechanical  properties of penta-SiC$_2$}

\begin{figure}
\centering
\includegraphics[width=0.75\linewidth]{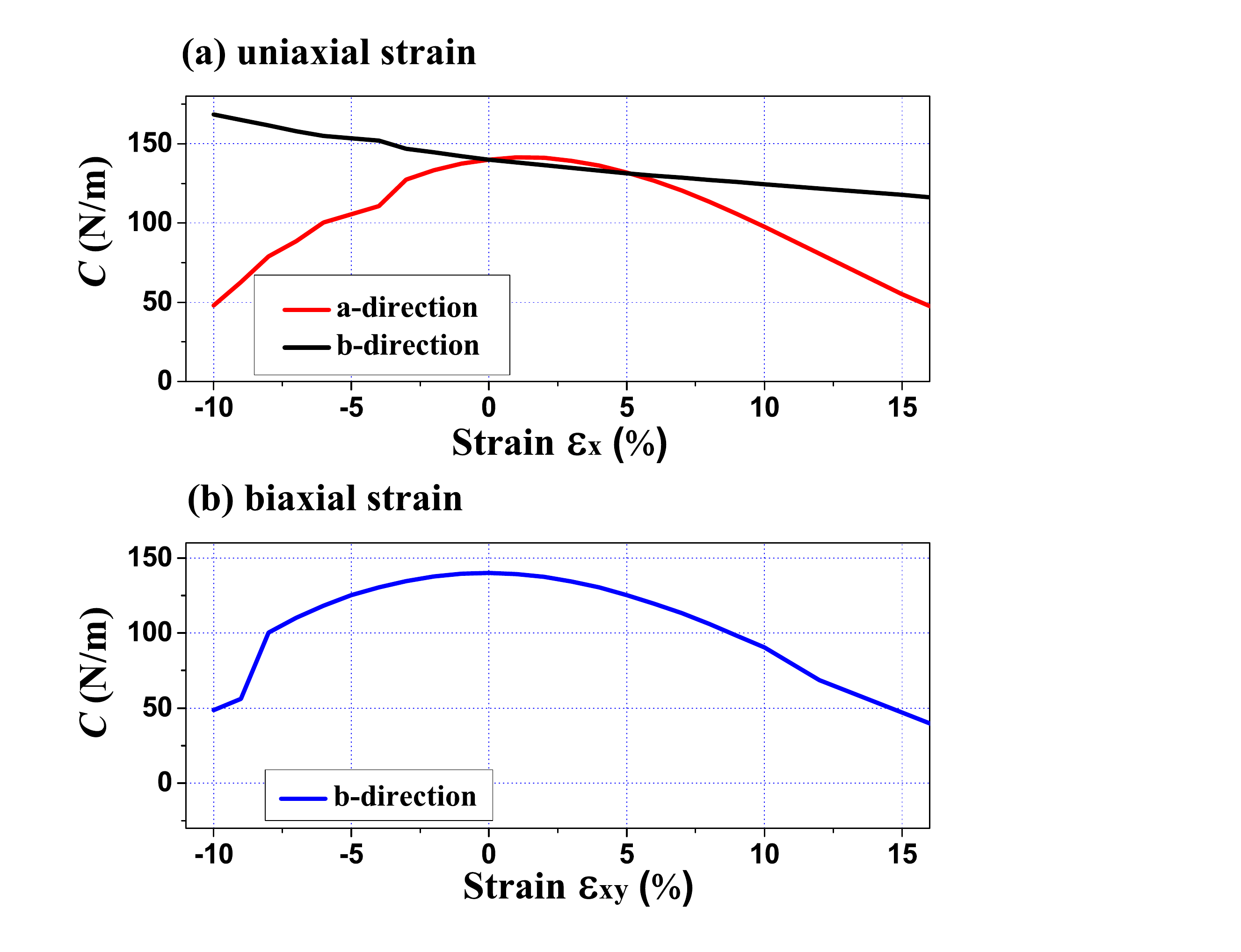}
\caption{Calculated elastic modulus $C_{ij}^{2D}$along a- and b-directions under various mechanical strain for penta-SiC$_2$.}
\label{elastic-sic} 
\end{figure}

We have studied the effects of in-plane uniaxial and biaxial strains on mechanical, electron and transport properties of penta-SiC$_2$. Here, $\varepsilon_x$, $\varepsilon_y$ and $\varepsilon_{xy}$ refer to the components of relative strain along $a-$, $b-$ and bi($a+b$)-directions respectively. Since penta-SiC$_2$ has nearly isotropic physical properties along a- and b-directions respectively, only the strain $\varepsilon_x$ is considered in the following. In addition, under the in-plane biaxial strain, the physical properties are nearly the same along a- and b-directions, so only the physical properties along the b-direction are presented here. The positive (negative) signs of strains represent tensile (compressive) strain, and the value of strain is evaluated as the lattice stretching (condensing) percentage. 

Due to three-dimensional (3D) periodic boundary conditions, the 2D elastic constant    $C_{ij}^{2D}$ should be rescaled by multiplying the $c$ lattice parameter corresponding to the vacuum space between 2D layers\cite{Blonsky2015}, $i.e.$, $C_{ij}^{2D}$= $c$ $C_{ij}^{3D}$. The calculated in-plane elastic constant $C_{11}^{2D}$ = $C_{22}^{2D}$=140 $N/m$ for penta-SiC$_2$, indicates that penta-SiC$_2$ has nearly isotropic mechanical property. The positive values of $C_{11}^{2D}$ and $C_{22}^{2D}$ mean that penta-SiC$_2$ is mechanically stable, according to the the mechanical stability criteria. The 3D Young's modulus $Y$ can be expressed as $Y$=$C_{ij}^{2D}$/$t$, where $t$ is the effective thickness of penta structure (4.8 $\AA$ for penta-SiC$_2$), and the calculated value of  $Y$ is 292 GPa for SiC$_2$, similar to that of MoS$_2$ (270 $\pm$ 100 GPa)\cite{Bertolazzi2011}, while larger than that of black phosphorene (experimental values: 179 GP and 55 GP along a- and b-directions respectively)\cite{Wei2014}, TiS$_3$ (96 GP and 153 GP along a- and b-directions respectively)\cite{kang2015} and HfS$_2$ (137 GPa)\cite{kang2015}. The difference can be attributed to the bond length and bond density of different 2D materials.  

Furthermore, we have calculated the elastic modulus along the a- and b-directions under various mechanical strain, as shown in Fig.~\ref{elastic-sic}. By stretching (compressiving) the atom-atom bond length, it is shown that a decreased (increased) of bond energy may reduce (enlarge) the value of elastic constant. Fig.~\ref{elastic-sic}(a) shows the dependence of elastic constant along a- and b-directions under uniaxial ($\varepsilon_x$) strain. Under uniaxial strain, the elastic constant decreases monotonously in the range of ${-10\%}\leq\varepsilon_x\leq{10\%}$ along a-direction, while along b-direction, it has a maximum value at a tensile strain of $\varepsilon_x$=1\% and decrease linearly from $\varepsilon_x$=1\% to $10\%$, and under compressive strain, the elastic constant decreases linearly at a high rate. While for biaxial strain as shown in Fig.~\ref{elastic-sic}(b), the value of elastic modulus of two directions decreases with similar tendency of b-direction under uniaxial strain for penta-SiC$_2$.

\subsection{Strain-engineering electronic properties of penta-SiC$_2$}

\begin{figure}
\centering
\includegraphics[width=0.75\linewidth]{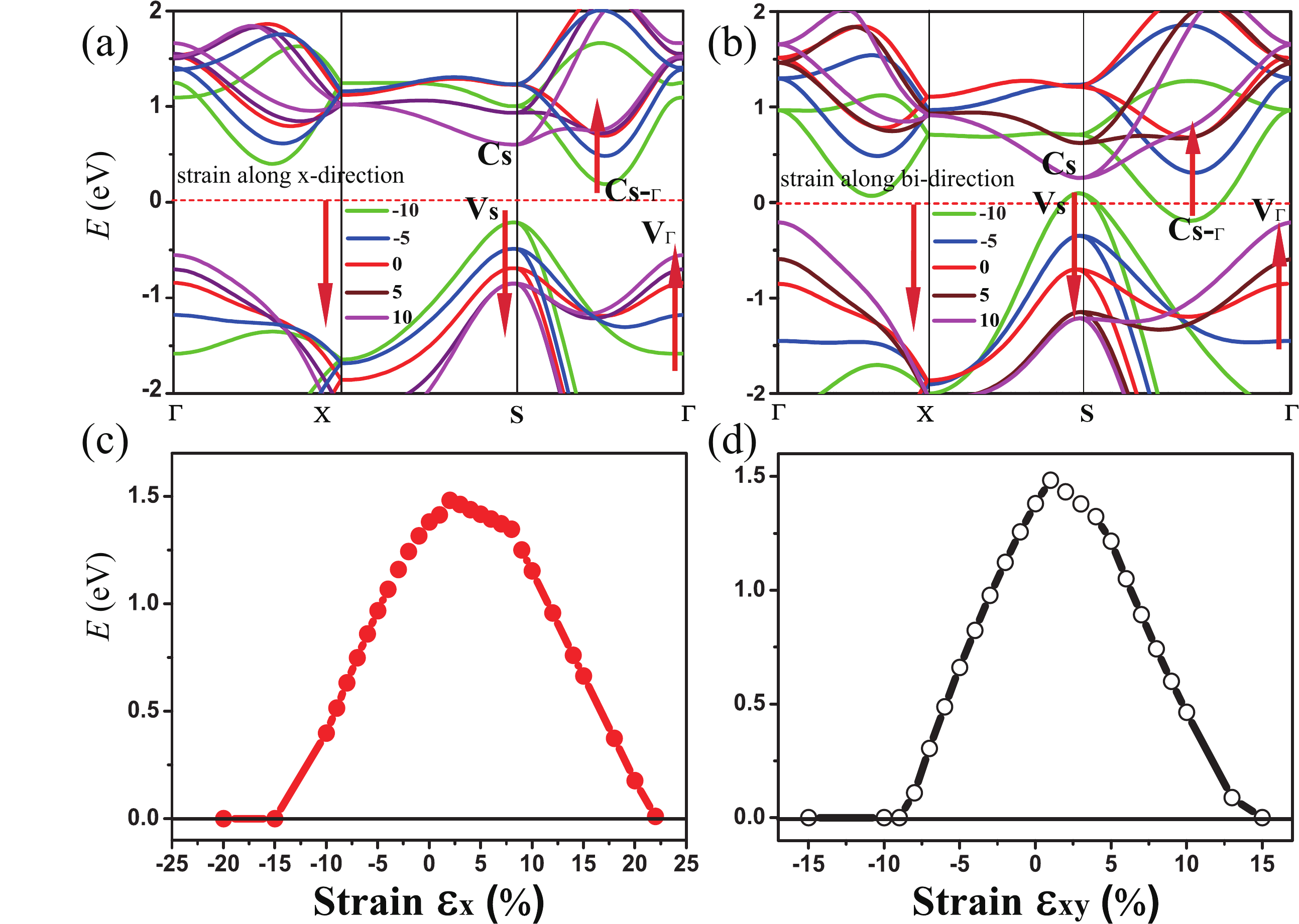}
\caption{PBE calculated band structures under (a) uniaxial ( a-direction) strain and (b) biaxial (bi-direction) strain for penta-SiC$_2$. (c) and (d) show the evolution of band gap for SiC$_2$ as a function of the applied various (uniaxial and biaxial) strain, respectively. Furthermore, the crystal structure is fully relaxed(both unit cell parameters and atomic positions under symmetry constraints).}
\label{band strain} 

\end{figure}

Fig.~\ref{band strain} shows the evolution of the valence and conduction band structures as a function of different strains, mainly from ${-10\%}$ to ${10\%}$ of the fully relaxed structure. Fig.~\ref{band strain} (a) and (b) show the dependence of the energy bands of monolayer penta-SiC$_2$ on strain along a- and bi-directions, respectively. The shift of the band edge as shown in Fig.~\ref{band strain} (a) and (b) can be understood in terms of the bonding and antibonding states\cite{PhysRevB.91.085423}. The VBMs ($V_s$ and $V_\Gamma$) and the CBMs ($C_s$ and $C_{s-\Gamma}$) are shown in Fig.~\ref{band strain} (a) and (b). The monolayer penta-SiC$_2$ is an indirect semiconductor described by VBM of $V_s$ and CBM of $C_{s-\Gamma}$. However, by applying an uniaxial tensile strain ($\varepsilon_x$), the energies corresponding to the $V_s$ and $C_{s-\Gamma}$ change greatly, which transforms penta-SiC$_2$ into a new type of indirect semiconductor formed by VBM of $V_\Gamma$ and CBM of $C_s$.  Under uniaxial compressive strain, $C_{s-\Gamma}$ decreases while $V_s$ increases, leading to a shrinking indirect band gap of the strained penta-SiC$_2$. While for an external strain along bi-direction ($\varepsilon_{xy}$), large changes take place for the energies corresponding to $V_s$, $V_\Gamma$, $C_s$ and $C_{s-\Gamma}$ as shown in Fig.~\ref{band strain} (b). When $\varepsilon_{xy}=-8\%$, the semiconductor(indirect)-to-metal transition takes place.  The strain tunable electronic structure of monolayer SiC$_2$ will significantly influence the electronic transport properties, which can be used in the electronic and optoelectronic applications. 

 Fig.~\ref{band strain} (c) and (d) shows the evolution of calculated band gap under various strains. When applying the in-plane uniaxial strain for ${-15\%}\leq\varepsilon_x\leq{22\%}$, penta-SiC$_2$ remains an indirect semiconductor with a maximum band gap of 1.481 eV at ${2\%}$, while the gap energies and the positions of the VBM and CBM change.  However, when applying the in-plane biaxial strain for ${-15\%}\leq\varepsilon_{xy}\leq{15\%}$,  penta-SiC$_2$ also remains the indirect semiconducting feature, and it converts into a metal beyond ${15\%}$ and ${-9\%}$.  However, the band gap reaches a maximum value of 1.483 eV at the strain of ${1\%}$.

As can be seen in Fig.~\ref{band strain} (c) and (d), the electronic band gap is very sensitive to the application of  compressive and tensile strain. The reason may be that  penta-SiC$_2$ exhibits a partial inversion of the vertical ordering of the p-p-$\sigma$ and p-p-$\pi$ electronic bands, and the energy states corresponding to the in-plane $\sigma$-orbitals are located in the vicinity of the Fermi level. Meanwhile, the in-plane $\sigma$-bonds is sensitive to variations in the bonding length, which is the main factor leading to the variation of  band gap upon different strains\cite{Littlewood2015}.

\begin{figure}
\centering
\includegraphics[width=0.75\linewidth]{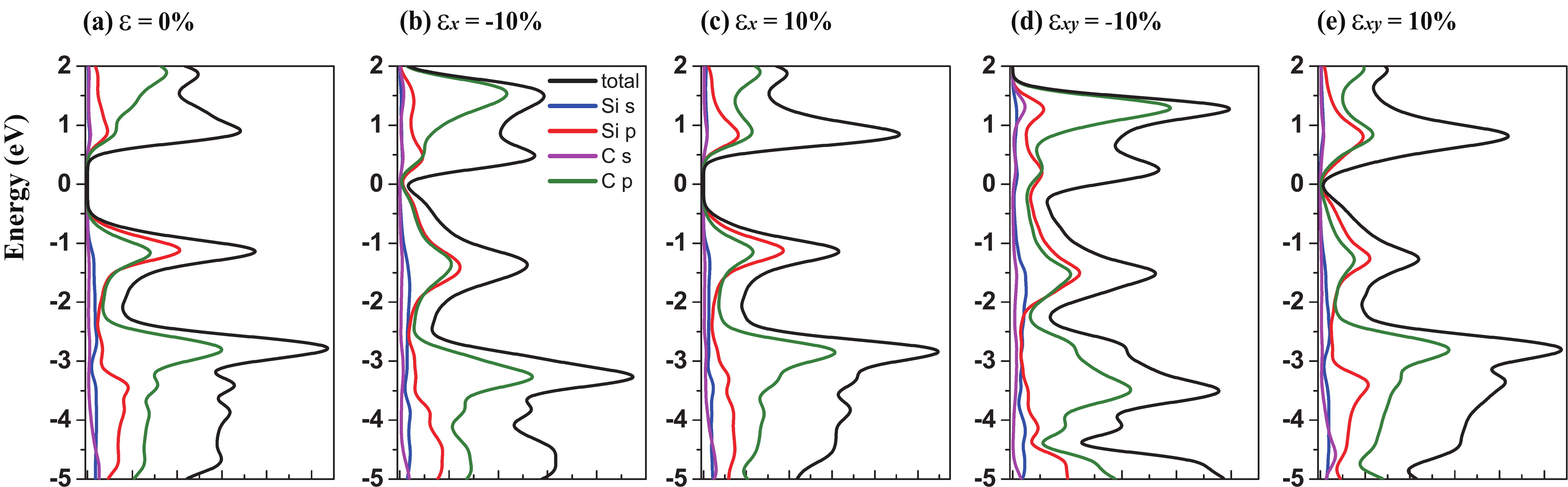}
\caption{PBE calculated total and partial densities of states for penta-SiC$_2$ under uniaxial and biaxial strain with compressive and tensile strength: (a) no strain, (b) $\varepsilon_x=-10\%$, (c) $\varepsilon_x=10\%$, (d) $\varepsilon_{xy}=-10\%$, and (e) $\varepsilon_{xy}=10\%$, respectively.}
\label{band dos} 
\end{figure}

In order to understand the contribution of different orbitals from different atoms to the electronic states, we have calculated the total (DOS) and partial (PDOS) densities of states for penta-SiC$_2$ with tensile and compressive strains: $\pm10\%$ strains along a-direction and bi-direction, respectively, as shown in Fig.~\ref{band dos}. 
By comparison, it's obvious that the contribution from different orbits of Si and C atoms to the valence band does not change too much when uniaxial or biaxial strain is applied. However, when uniaxial or biaxial compressive strain larger than $\varepsilon_x=-10\%$ is applied, the contribution from C 2$p$ states becomes dominant. Therefore, when monolayer penta-SiC$_2$is free of strain or under small strain (less than $10\%$), the conduction band mainly constructs the low hybridization states of Si 3$p$ and C 2$p$ orbits, when a large strain is applied, the contribution from C 2$p$ states overwhelms that from Si 3$p$ states.

\subsection{Strain-engineering transport properties of penta-SiC$_2$}
Here we perform the deformation potential theory, which was proposed by Bardeen and Shockley\cite{deformation1950} to analyze the intrinsic carrier mobility $u$, and  has been widely used in low-dimensional systems\cite{Shuai2011, Shuai2013, Cai2014, Wang2015b}. Based on the acoustic phonon limited approach\cite{deformation1950}, the charge carrier mobility of 2D material can be written as,
\begin{equation}\label{mobilities}
\mu = \frac{2e\hbar^3C}{3k_BT|m^*| ^2E_l^2},
\end{equation}
where $e$ is the electron charge, $T$ is the temperature which is equal to 300 K throughout the paper. $C$ is the elastic modulus of a uniformly deformed crystal by strains, $m^*$ is the effective mass given by $m^*=\hbar^2({\partial^2{E(k)}/{\partial{k^2}})}^{-1}$ (where $\hbar$ is the reduced Planck's constant, $k$ is the magnitude of wave-vector in momentum space, and $E(k)$ denotes the energy corresponding to the wave vector). $E_l$ is the DP constant defined by $E_l^{e(h)}=\Delta{E_{CBM(VBM)}}/(\delta{l}/l)$, where $\Delta{E_{CBM(VBM)}}$ is the energy shift of the band edge with respect to the vacuum level under a small dilation $\delta{l}$ of the lattice constant $l$. 

\begin{figure}
\centering
\includegraphics[width=0.75\linewidth]{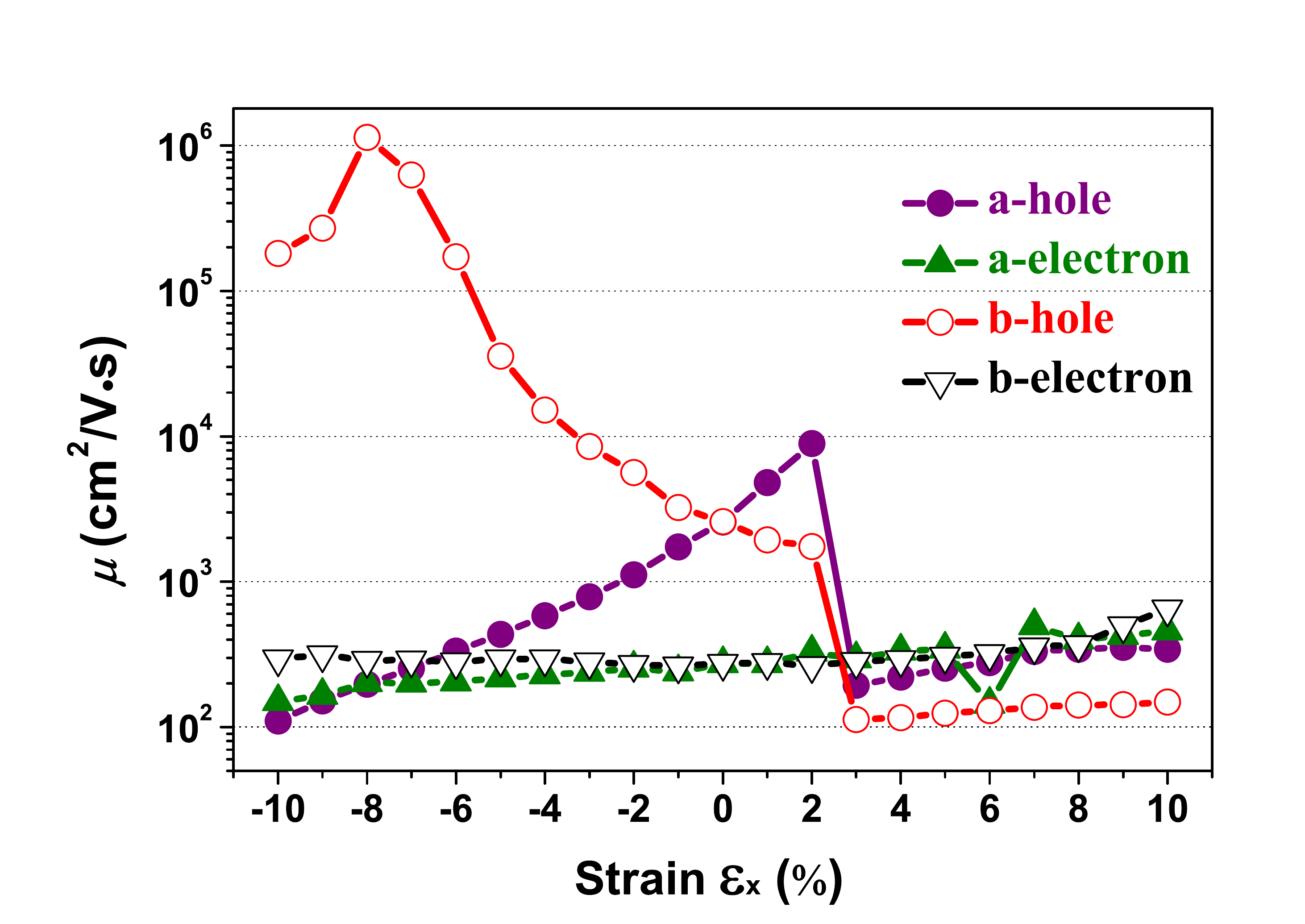}
\caption{Calculated charge carrier (hole and electron) mobility ($u$) along the a- and b-directions as a function of the applied uniaxial  strain, respectively.}
\label{mobility} 
\end{figure}

According to Eq. (\ref{mobilities}), the carrier mobility of electrons and holes along certain directions for penta-SiC$_2$ are obtained, as shown in Fig.~\ref{mobility} and Table.~\ref{mobility-bi}. Although the PBE calculations always underestimate the band gap, the calculated carrier mobilities are in good agreement with experimental results for many 2D materials\cite{Dai2015, Cai2014, Shuai2013, Shuai2012, Shuai2011}. Firstly, we compare the a-direction uniaxial strain modulated charge carrier mobility of penta-SiC$_2$ along a- and b-directions as shown in Fig.~\ref{mobility}. Three different physical parameters, namely carrier effective mass ($m^*$), DP constant ($E_l$) and elastic modulus ($C$), are subjected to change at a particular temperature under strain.

\begin{figure}
\centering
\includegraphics[width=0.75\linewidth]{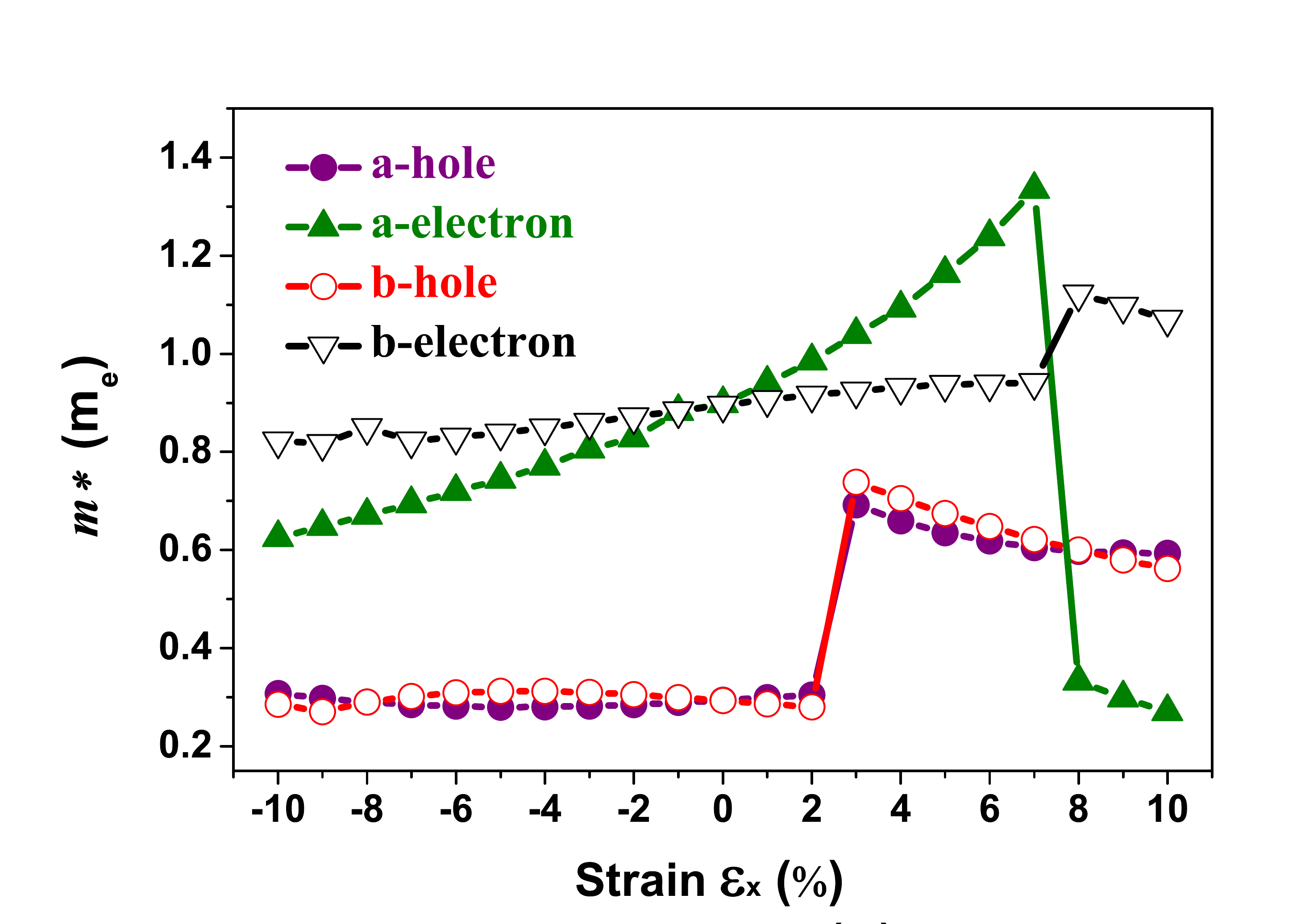}
\caption{Respective dependence of calculated carrier (hole $m_h^*$ and electron $m_e^*$) effective masses  along the a- and b-directions on the applied uniaxial strains.}
\label{effmass} 
\end{figure}

The band structures of strained penta-SiC$_2$ change when the structure changes, which subsequently alter the effective mass of carriers determined by the curvatures of  band edges near the Fermi level. Fig.~\ref{effmass} demonstrates the evolution of the calculated effective masses of electrons ($m_e^*$) and holes ($m_h^*$) along a- and b-directions under uniaxial strains along a-direction. The $m_h^*$ at $S$ point is 0.29$m_0$ and  $m_e^*$ at $S-\Gamma$ point is 0.90$m_0$ without strain ($m_0$ is the static electron mass).  As shown in Fig. \ref{band strain}(a), applied strains along a-direction $\varepsilon_x$ may lead to large shifts of the positions of VBMs and CBMs in the band structures. The VBM of penta-SiC$_2$ shifts from $V_s$ to $V_\Gamma$ with the strain of $\varepsilon_x=3\%$, which results in a dramatic increase of $m_h^*$ ($0.30m_0\rightarrow0.69m_0$ along a-direction, and $0.28m_0\rightarrow 0.74m_0$ along b-direction) when $\varepsilon_x$ increases from $2\%$ to $3\%$. Meanwhile, The CBM of penta-SiC$_2$ shifts from $C_{s-\Gamma}$ to $C_s$ with the strain of $\varepsilon_x=8\%$. Therefore a dramatic change of electron effective mass occurs as well when the uniaxial strain changes from $\varepsilon_x=7\%$ to $\varepsilon_x=8\%$. The subsequent $m_e^*$ reduces significantly from 1.33$m_0$ to 0.33$m_0$, and the b-direction $m_e^*$ increases from 0.94$m_0$ to 1.12$m_0$, as shown in Fig. \ref{effmass}. Additionally, the anisotropic effective mass under strains will lead to anisotropic carrier mobilities, which finally lead to direction-dependent electron conductivity. 

\begin{figure}
\centering
\includegraphics[width=0.75\linewidth]{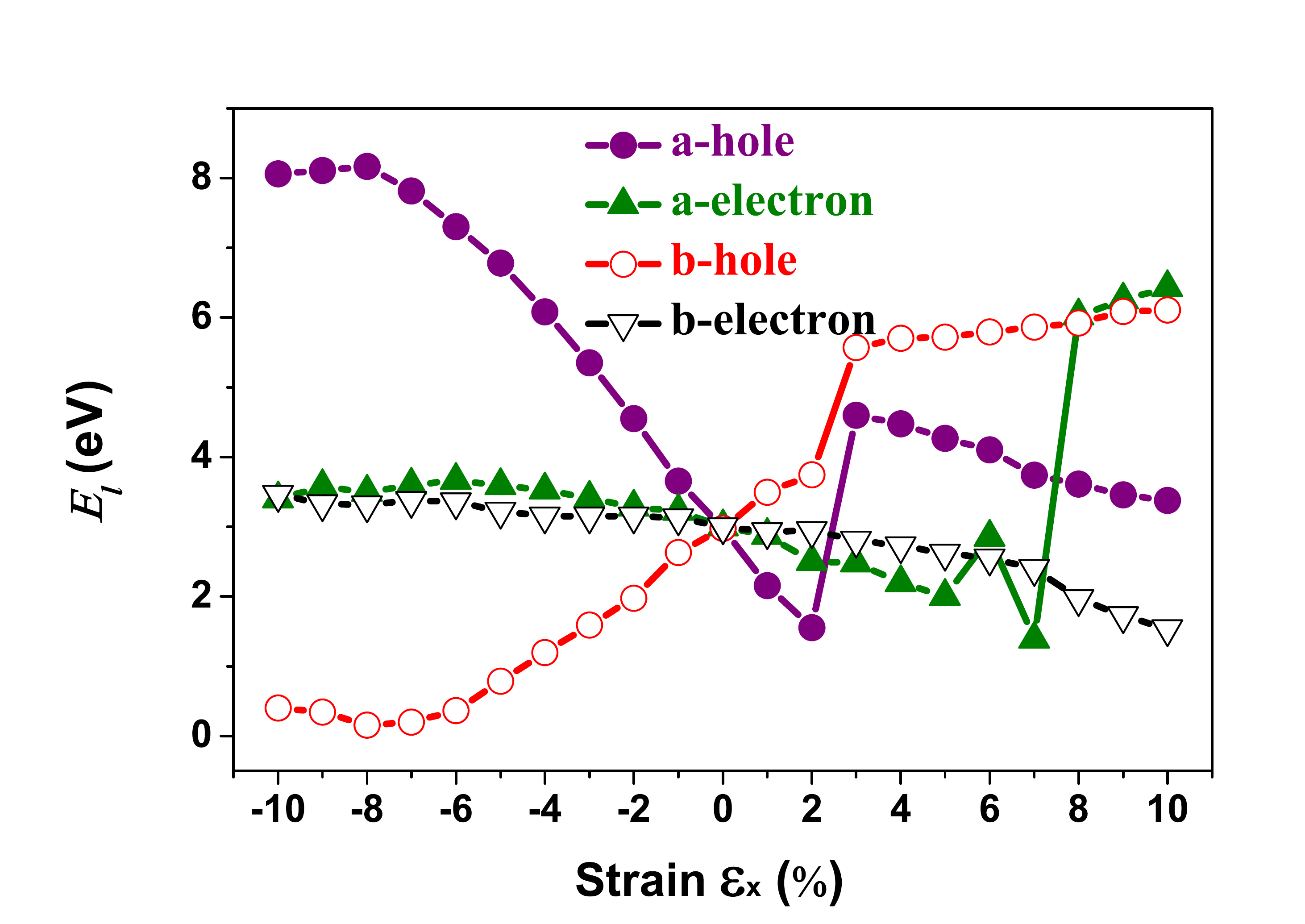}
\caption{Calculated carrier (hole and electron) deformation potential constant ($E_l$) along the a- and b-directions as a function of the applied uniaxial  strain.}
\label{deformation} 
\end{figure}

In order to calculate DP constant, a small dilation need to be applied along the direction. The calculated DP constant for holes is 2.98 eV and 3.00 eV for electrons, respectively, which are in good  agreement with previous theoretical results\cite{Li2015}. Fig.~\ref{deformation} demonstrates the evolution of the electron ($E_{l-e}$) and hole ($E_{l-h}$) DP constants along a- and b-directions under uniaxial strains along a-direction. Similar to the case of effective masses, the shifts of VBMs with a uniaxial strain of $\varepsilon_x=3\%$  and CBMs with a strain of $\varepsilon_x=8\%$ cause dramatic changes of DP constants of holes and electrons, respectively, as shown in Fig.~\ref{deformation}.  Interestingly, the obtained b- direction $E_{l-h}$ of holes is low ($0.15 eV\rightarrow0.40 eV$) in the strain range of ${-10\%}\leq\varepsilon_x\leq{-6\%}$, which subsequently results in high hole-carrier mobilities larger than $10^5 cm^2/V s$ along a particular direction as shown in Fig.~\ref{mobility}.

The dependence of calculated carrier mobilities on the uniaxial strain is shown in Fig.~\ref{mobility}. The obtained hole and electron mobilities without strain are 2.59$\times10^3 cm^2/V s$ and 2.74 $\times10^2 cm^2/V s$, respectively. For uniaxial strain along the a-direction, dramatic enhancement of the hole mobility along the b-direction under compressive strains is observed, increasing from 2.59$\times10^3 cm^2/V s$ to 1.14 $\times10^6 cm^2/V s$ at $\varepsilon_x=-8\%$. In Table.~\ref{table-2}, the calculated carrier mobilities of strained penta-SiC and some widely-investigated two-dimensional semiconductor are listed. By comparison, it is found that, the hole carrier mobility of penta-SiC$_2$ stained by $\varepsilon_x=-8\%$ is ultrahigh, an order of magnitude larger than those of graphene (2.58 $\times10^5 cm^2/V s$)\cite{shao2013}, silicene (3.39 $\times10^5 cm^2/V s$)\cite{shao2013}, three or four orders higher than those of monolayer black phosphorene (2.2 $\times10^3 cm^2/V s$)\cite{Fei2014}, MoS$_2$($\sim200 cm^2/V s$)\cite{Cai2014,Kim2012}, which indicates that strain-engineering monolayer pent-SiC$_2$ can be quite promising for the application of microelectronics.

Meanwhile, strain-engineering has little influence on the electron mobility along both a- and b-directions, as shown in Fig.~\ref{mobility}. The hole mobility along the a-direction increases nearly linearly from 2.59$\times10^3 cm^2/V s$ to 1.0 $\times10^4 cm^2/V s$ in the strain range of $\varepsilon_x=-10\%$ to $2\%$, which can be attributed to the monotonic decrease of effective masses of holes along a-direction, according to Eq. (\ref{mobilities}).

\begin{table}
\centering
\caption{Calculated carrier mobility of strained pengta-SiC in comparison to those of monolayer graphene, silicene, balck phosphorene and MoS$_2$. (Unit of carrier mobilities, $cm^2/V\cdot {s}$.)}
\begin{tabular}{cccccc}
\hline
    carrier mobility & penta-SiC$_2$ &  graphene\cite{shao2013} &  silicene\cite{shao2013} &  balck phosphorene \cite{jun2014,Fei2014}& MoS$_2$\cite{Cai2014,Kim2012} \\
\hline
hole & 1.14 $\times10^6 $  & 3.51 $\times10^5$  & 2.23 $\times10^5 $& 286 & 200  \\
\hline
electron& 649  & 3.39 $\times10^5 $ & 2.58 $\times10^5 $ & 2.2$\times10^3 $ & 72, $\sim 200$\\
\hline
\end{tabular}
\label{table-2}
\end{table}

\begin{table}
\centering
\caption{The effective mass and charge carrier mobilities in penta-SiC$_2$, It shows the calculated effective mass $m_b^*$ (with $m_0^*$ being the static electron mass), deformation potential constant $E_{l-b}$, 2D elastic modulus $C_b$ along the $\Gamma-Y$ direction, The electron and hole carrier mobility ($\mu_b$) are calculated by using Eq.(1) at $T$=300 K. }
\begin{tabular}{cccccccccccc}
\hline
 Carrier type &  $\varepsilon_{xy}$  &  $m_b^*$ &  $E_{l-b}$ &  $C_b$  & $\mu_b$ &Carrier type &  $\varepsilon_{xy}$ &  $m_b^*$  & $E_{l-b}$ &  $C_b$ &
 $\mu_b$\\
 &[\% ] & [$m_0$] & [$eV$] &  [$N/m$]&  [$\times10^2 cm^2/V\cdot {s}$] & & [\% ]  &[$m_0$] &  [$eV$]&[$N/m$] &[$\times10^2 cm^2/V\cdot {s}$] \\
\hline
hole & -10 & 0 &0&0&0&electron & -10 & 0 &0 & 0 &  0 \\
  & -8& 0.29 &5.45&100.37&5.78& & -8 & 0.67 &3.66 & 100.37 & 2.24 \\
  & -6 & 0.30 &4.78&118.29&8.43& & -6 & 0.71 &3.55 & 118.29 &  2.60 \\
  & -4 & 0.30 &4.12&130.41&12.31& & -4 & 0.75&3.47 & 130.41 &  2.72 \\
  & -2 & 0.30&4.03&137.66&13.68& & -2 & 0.80 &3.31 & 137.66 &  2.76 \\
  & 0 & 0.29 &2.99&139.89&25.88& & 0 & 0.90 &3.0 & 139.89 &  2.74\\
  & 2 & 0.66 &5.29& 137.31&1.58& & 2 & 1.05 &2.61 & 137.31 &  2.60\\
  & 4 & 0.55 &4.80&130.34&2.65& & 4 & 1.30 &1.81 & 130.34 &  3.30\\
  & 6 & 0.48 &4.34&119.47&3.91& & 6 & 0.52 &4.13& 119.47 &  3.62 \\
  & 8 & 0.43&4.00&106.01&4.99& & 8 & 0.46 &3.96& 106.01 &  4.44 \\
  & 10 & 0.40 &3.62&90.39&6.03& & 10 & 0.43 &3.68 & 90.39 &  5.14 \\
\hline
\end{tabular}
\label{mobility-bi}
\end{table}

Table.~\ref{mobility-bi} shows the evolution of the effective mass ($m_b^*$), DP constant ($E_{l-b}$), elastic modulus ($C_b$) and the calculated results of hole and electron mobilities along b-direction under biaxial strains. For the hole mobility as listed in Table.~\ref{mobility-bi}, since all the three decisive parameters ($m_b^*$, $E_{l-b}$ and $C_b$) increase by applying tensile and compressive strain, the hole mobilities for strained structures are thus smaller than that without strain according to Eq. (\ref{mobilities}). The obtained electron mobility increases monotonically from $\varepsilon_{xy}=-10\%$ to $\varepsilon_{xy}=10\%$, mainly caused by the decrease of the effective mass $m_e^*$.

\section{Conclusion}
To summarize, we have demonstrated the evolution of structure, electronic and charge carrier transport properties of penta-SiC$_2$ under uniaxial or biaxial strain. Based on the first-principles calculation using the PBE functional method, SiC$_2$ is found to be a semiconductor with a direct band gap of 1.39 eV,  and HSE06 functional gives the band gap of 2.85 eV. By employing Voigt notation, penta-SiC$_2$ has a relatively high Young's modulus: 292 GPa. The transition point from semiconductor  to metal is different between uniaxial and biaxial strains. The hole mobility along b-direction can be significantly improved by a-direction compressive strain. Meanwhile, strain-engineering has relatively little effect on electron mobility. Therefore, it would be helpful to select carrier type based on their different transport properties under various strains. This study gives us a comprehensive understanding of the strain effect on the electronic properties of monolayer penta-SiC$_2$, and the properties may find applications in nanoscale strain tensor and conductance-switch field effect transistor storages.

\section*{Acknowledgement}
This work is supported by the National Natural Science Foundation of China under Grants No. 11374063 and 11404348, and the National Basic Research Program of China (973 Program) under Grants No. 2013CBA01505.


\end{document}